# Degradation-Aware Lifecycle Planning of Microgrids with Second-Life Batteries through Reliability-Guided Refinement

Hassan Zahid Butt and Xingpeng Li

***Abstract*— Second-life batteries (SLBs) can reduce storage investment in microgrids, but their lower initial state of health and round-trip efficiency (RTE) can make degradation-naive sizing unreliable over long planning horizons. This paper develops DAVIR-MG, a degradation-aware validation and investment refinement framework for microgrid planning. A 20-year mixed-integer linear planning model determines installed capacities of dispatchable generation (DG), photovoltaic (PV) generation, and a battery energy storage system (BESS) using 365 days of hourly data. The resulting portfolio is assessed through sequential lifecycle validation that accounts for PV degradation, battery capacity fade, RTE decline, and replacement. When validation identifies degradation-induced energy not served (ENS) above the prescribed tolerance, technology-specific refinement options are evaluated and compared using validated lifecycle cost. The framework is evaluated across tariff structures, PV degradation rates, new-BESS and SLB cases, degradation-mechanism ablation, technology-specific reliability refinements, Cost-ENS trade-offs, and SLB break-even conditions. Across seven cases, validated net present cost (NPC) exceeded planning NPC by 3.0%–11.6%, while adequacy shortfalls emerged only after year 13. In the selected SLB case, capacity fade alone accounted for 76.4% of cumulative ENS, RTE decline alone for 10.7%, and their interaction for the remaining 12.9%. BESS and DG refinements restored adequacy at nearly identical lifecycle cost but through different technical pathways. Under equal-reliability sizing, the minimum SLB discount required for cost parity increased from 22.9% to 35.7% when future replacement was priced at the new-BESS cost. DAVIR-MG provides a practical, industry-relevant, and computationally manageable framework for identifying degradation-induced adequacy risks and supporting informed lifecycle investment decisions.**

## NOMENCLATURE

Sets and indices:

| | |
|---|---|
| $Y$ | Planning years; index $y$ |
| $D$ | Operating days; index $d$ |
| $T$ | Hourly periods; index $t$ |

Parameters:

| | |
|---|---|
| $P_{y,d,t}^{L}$ | Electrical demand (MW) |
| $P_{d,t}^{PV,cf}$ | PV capacity factor (p.u.) |
| $\phi_y^{PV}$ | PV performance factor in year $y$ |
| $H_y$ | BESS state of health used in year $y$ |
| $\eta_y^{RTE}$ | BESS round-trip efficiency |
| $\eta_y^{ch}, \eta_y^{dch}$ | BESS charging and discharging efficiencies |
| $q_y$ | Discount factor for year $y$ |
| $\Delta t$ | Duration of one operating interval (h) |
| $C_{DG}^{op}$ | DG variable operating cost ($/MWh) |
| $C_{DG}^{nl}$ | DG no-load operating cost ($/h) |
| $C_x^{cap}$ | Capital cost of technology $x$, $x \in \{DG, PV, B\}$ |
| $C_x^{fom}$ | Fixed O&M cost of technology $x$, $x \in \{DG, PV\}$ |
| $C_{y,d,t}^{imp}, C_{y,d,t}^{exp}$ | Grid import and export prices ($/MWh) |
| $C^{LS}$ | Load-shedding penalty cost ($/MWh) |
| $P^{tie}$ | Grid tie-line capacity (MW) |
| $P^{DG,min}$ | Minimum DG output (MW) |
| $SOC^{min}, SOC^{max}$ | Minimum and maximum BESS SOC |
| $SOC^{init}$ | Initial normalized BESS SOC |
| $\tau^{ch}, \tau^{dch}$ | BESS charge and discharge durations (h) |
| $M^{DG}, M^{B}$ | DG and BESS upper-bound constants |

Variables:

| | |
|---|---|
| $S^{DG}$ | Installed DG capacity (MW) |
| $S^{PV}$ | Installed PV capacity (MW) |
| $S^{B}$ | Installed BESS nameplate capacity (MWh) |
| $P_{y,d,t}^{DG}$ | DG output (MW) |
| $P_{y,d,t}^{ch}, P_{y,d,t}^{dch}$ | BESS charging and discharging power (MW) |
| $P_{y,d,t}^{imp}, P_{y,d,t}^{exp}$ | Grid import and export power (MW) |
| $P_{y,d,t}^{PV,curt}$ | Curtailed PV power (MW) |
| $P_{y,d,t}^{LS}$ | Unserved load (MW) |
| $E_{y,d,t}^{B}$ | Energy stored in the BESS (MWh) |
| $u_{y,d,t}^{DG}$ | Binary DG commitment variable |
| $u_{y,d,t}^{ch}, u_{y,d,t}^{dch}$ | Binary BESS charging and discharging states |
| $u_{y,d,t}^{imp}, u_{y,d,t}^{exp}$ | Binary grid import and export states |

## I. INTRODUCTION

Microgrids (MGs) coordinate distributed generation, renewable energy resources, energy storage, and local demand within an integrated energy system [1]–[3]. Their ability to use local resources with varying levels of grid support can improve resilience and reduce dependence on external electricity or fuel supplies, particularly in grid-constrained and remote applications [4], [5]. Solar photovoltaic (PV) generation has become a prominent MG resource because of its broad deployment potential and compatibility with distributed energy systems [6]. However, its variable output requires flexible resources that can balance generation and demand across changing operating conditions.

Battery energy storage systems (BESSs) provide this flexibility through energy shifting, peak-demand support, grid-interaction management, and backup supply [7], [8]. Their

Hassan Zahid Butt and Xingpeng Li are with the Department of Electrical and Computer Engineering, University of Houston, Houston, TX, 77204, USA. (e-mail: hbutt@uh.edu; xli82@uh.edu).

economic and operational value depends on the installed energy capacity, utilization, efficiency, and service life. An undersized BESS may fail to maintain adequate supply as demand grows and battery performance declines, whereas an oversized system increases initial investment and may remain underused. Long-term MG planning must therefore consider both the initial investment decision and the evolution of asset performance over the project horizon.

PV systems and BESSs both experience performance degradation. PV output gradually declines because of material aging and environmental exposure, while lithium-ion batteries undergo capacity and efficiency losses that depend on their operating conditions and cycling history [9]–[12]. Capacity fade reduces the energy available for dispatch, and round-trip efficiency (RTE) decline raises charging and discharging losses, together influencing the ability to meet demand during constrained periods. A model that holds BESS capacity and RTE at their initial values may therefore underestimate lifecycle cost and conceal shortfalls that emerge late in life.

The growing retirement of electric vehicle batteries creates an opportunity to reduce stationary storage costs through battery repurposing [13]–[15]. Cells that no longer meet automotive requirements may retain sufficient capacity for stationary duty, and using them as second-life batteries (SLBs) extends their useful life before recycling [16], [17]. However, SLBs enter stationary service with lower state of health (SOH), lower RTE, and possibly shorter remaining life. Their initial cost advantage must therefore be evaluated against the additional capacity, operating losses, and replacement expenditure required to maintain the same level of service. Comparisons based only on installed battery cost can consequently lead to misleading investment conclusions. A low-cost SLB may appear attractive in the initial planning solution but become less competitive after degradation, reliability, and replacement requirements are considered. A meaningful comparison should evaluate new BESSs and SLBs under a common lifecycle-reliability criterion and account for both initial and future replacement costs [18]–[20].

Representing these effects within long-horizon capacity planning is computationally challenging. Battery degradation strongly depends on the depth of charge-discharge cycles, which couples hourly dispatch with SOH, RTE, available capacity, and replacement timing [21]–[23]. Electrochemical and semi-empirical models may improve fidelity, but their inclusion in long-horizon mixed-integer linear programming (MILP) models can produce substantial computational complexity. Simplified throughput penalties or fixed annual degradation rates do not capture the relationship between optimized operation and subsequent battery condition.

This paper develops DAVIR-MG, a degradation-aware validation and investment refinement framework for lifecycle MG planning. The framework combines multi-year MILP-based capacity planning with sequential lifecycle validation to evaluate the effects of PV degradation, BESS capacity fade, RTE decline, and battery replacement. When validation identifies degradation-induced energy not served (ENS), technology-specific refinements are assessed under a common reliability criterion and compared using validated economic net present cost (NPC). The framework is further used to isolate the reliability effects of different degradation mechanisms, quantify the cost-reliability trade-off of BESS augmentation, and compare new batteries and SLBs after both are sized to satisfy the same lifecycle-reliability requirement.

## II. Literature Review

Long-term MG planning typically co-optimizes investment and operation of resources to minimize lifecycle cost subject to operational and reliability requirements [24]-[27]. MILP formulations are common because they represent discrete investment decisions, chronological dispatch, grid transactions, and operating-state logic. Full multi-year hourly models are heavy, so many studies adopt representative days or aggregated periods [28], [29], which improves tractability but can suppress the short-duration conditions that drive storage cycling, tariff response, and resource adequacy.

Degradation has been incorporated into MG planning at varying levels of detail. Cardoso et al. embedded linear battery aging in a multi-energy MG design model [23], Amini et al. treated sizing, degradation, service life, and replacement timing jointly [30], and Fallahifar and Kalantar represented depth-dependent capacity loss in storage planning [31]. Degradation assumptions clearly affect storage investment, although simplified capacity-loss relationships are generally required for tractability. Chronological approaches tie battery operation more closely to lifecycle condition [32]. Rainflow counting extracts partial cycles and depths from SOC trajectories so damage can be evaluated against cycle-life data [21], [33]. Fioriti et al. applied a rainflow-based model to multi-year residential BESS sizing [34], and Seger et al. used sequential SOH updates to integrate degradation and replacement into isolated MG design [35]. Both improve lifecycle representation but add computational burden when coupled directly to investment optimization.

Most degradation-aware planning focuses on capacity fade and assumes constant efficiency, although experimental and field studies show that RTE also declines with age and operating conditions [10]–[12], [16]. This is particularly relevant for SLBs, which enter stationary service already degraded, while the separate and combined effects of capacity and RTE loss on reliability, dispatch, and replacement remain insufficiently quantified. PV degradation, typically represented by annual derating [9], [36], further affects investment through its interaction with PV capacity, load growth, tariffs, DG operation, and battery condition. Few studies consider PV degradation and usage-dependent battery aging together at full chronological resolution.

SLBs have been examined extensively as a lower-cost alternative to new stationary batteries, spanning technical suitability, remaining life, repurposing, efficiency, and environmental benefit [13]-[17]. Techno-economic analyses show that competitiveness depends on acquisition cost, usable capacity, efficiency, service life, and replacement assumptions [18]-[20], so comparisons at equal nameplate capacity or equal initial cost mislead when the two technologies need different capacities for equivalent lifecycle reliability. Bai et al. proposed a reliability-oriented iterative design framework that separates planning from degradation validation for isolated MGs with SLBs [37], demonstrating the value of iterative lifecycle assessment and price-ceiling analysis. However, their scope centers on an isolated renewable MG with storage-focused refinement, leaving open the degradation-mechanism decomposition including interaction, cross-technology refinement, conditional Cost-ENS analysis, and equal-reliability break-even under alternative replacement pricing.

### *A. Research Gaps and Contributions*

Detailed degradation models remain difficult to embed in long-horizon hourly planning, while simplified formulations can conceal shortfalls that appear late in the project life. Existing work concentrates on capacity fade, with limited attention to the separate and interacting effects of RTE decline. Iterative approaches generally refine storage alone and rarely compare BESS, DG, and PV augmentation under a common criterion, and reliability is often treated as binary, giving little insight into the cost of progressively reducing degradation-induced ENS. For SLBs, economic assessments commonly emphasize acquisition or levelized storage cost without first sizing both technologies to equivalent lifecycle reliability. These gaps motivate a framework linking investment planning, degradation-aware validation, reliability refinement, and equal-service economic comparison.

The main contributions of this study are:

i. DAVIR-MG links a long-term MILP capacity-planning model with sequential degradation-aware validation using 365 days of hourly data.
ii. It quantifies the individual and interacting effects of BESS capacity fade and RTE decline on ENS, replacement timing, and lifecycle performance.
iii. BESS, DG, and PV refinements are evaluated under a common reliability criterion, and a conditional Cost-ENS analysis quantifies the trade-off between partial and complete reliability restoration.
iv. New and second-life batteries are separately sized to satisfy the same lifecycle-reliability requirement, followed by an SLB break-even assessment under alternative replacement-pricing conventions.

## III. Proposed Methodology

DAVIR-MG is a three-stage framework that integrates long-term investment planning, sequential lifecycle validation, and reliability-guided refinement, as illustrated in Fig. 1. It preserves chronological hourly operation while avoiding the computational burden of embedding all usage-dependent degradation and replacements within a single multi-year optimization model. In Stage 1, a multi-year MILP determines the installed DG, PV, and BESS capacities while accounting for load growth, PV degradation, grid transactions, capital investment, fixed O&M, and hourly operating costs. BESS SOH and RTE are held at their initial values during this stage, yielding the BESS-degradation-naive planning portfolio.

Stage 2 fixes that portfolio and solves sequential annual operating models. Each year updates load level, PV performance, BESS SOH, RTE, and usable energy capacity; the annual SOC trajectory is then processed by rainflow cycle counting and Miner's cumulative-damage rule to estimate cycling-induced degradation [33], [38], with replacement scheduled once the operational end-of-life (EOL) threshold is reached. If the validated portfolio fails either lifecycle-reliability criterion, Stage 3 evaluates BESS, DG, and PV refinements separately while holding the remaining capacities fixed, comparing lifecycle-feasible alternatives on validated economic NPC excluding the reliability penalty. The selected refinement is the least-cost feasible option among those evaluated.

### *A. Multi-Year Investment Planning Model*

The planning horizon comprises years $y \in Y$, days $d \in D$, and hours $t \in T$. The investment variables $S^{DG}$, $S^{PV}$, and $S^{B}$ denote the installed DG capacity in MW, PV capacity in MW, and BESS nameplate energy capacity in MWh, respectively. The annual discount factor, load growth, and PV performance are represented as

$$q_y = \frac{1}{(1+r)^{y-1}} \tag{1}$$

$$P^{L}_{y,d,t} = P^{L}_{1,d,t}(1+g_L)^{y-1} \tag{2}$$

$$\phi^{PV}_{y} = \phi^{PV}_{1}(1-\delta^{PV})^{y-1} \tag{3}$$

where $r$, $g_L$, and $\delta^{PV}$ are the real discount rate, annual load-growth rate, and annual PV degradation rate, respectively. The planning objective is given by (4).

$$\begin{aligned} \min\ & C^{cap}_{DG}S^{DG} + C^{cap}_{PV}S^{PV} + C^{cap}_{B}S^{B} + \\ & \textstyle\sum_{y\in Y} q_y [C^{fom}_{DG}S^{DG} + C^{fom}_{PV}S^{PV} + \\ & \textstyle\sum_{d\in D}\sum_{t\in T}(C^{op}_{DG}P^{DG}_{y,d,t} + C^{nl}_{DG}u^{DG}_{y,d,t} + C^{imp}_{y,d,t}P^{imp}_{y,d,t} - \\ & C^{exp}_{y,d,t}P^{exp}_{y,d,t} + C^{LS}P^{LS}_{y,d,t})\,\Delta t] \end{aligned} \tag{4}$$

Equation (4) minimizes initial DG, PV, and BESS capital cost plus discounted fixed O&M, DG operating and no-load cost, grid purchases, export revenue, and the load-shedding penalty. $C^{LS}$ is set high enough to prioritize demand service but is excluded from validated economic NPC because it is a model-enforcement term. Degradation and replacement costs are added only during lifecycle validation.

$$\begin{aligned} & P^{DG}_{y,d,t} + P^{dch}_{y,d,t} + \phi^{PV}_{y}P^{PV,cf}_{d,t}S^{PV} + P^{LS}_{y,d,t} + P^{imp}_{y,d,t} \\ & = P^{L}_{y,d,t} + P^{ch}_{y,d,t} + P^{PV,curt}_{y,d,t} + P^{exp}_{y,d,t} \end{aligned} \tag{5}$$

$$0 \le P^{PV,curt}_{y,d,t} \le \phi^{PV}_{y}P^{PV,cf}_{d,t}S^{PV} \tag{6}$$

$$0 \le P^{LS}_{y,d,t} \le P^{L}_{y,d,t} \tag{7}$$

Equation (5) enforces hourly supply-demand balance between generation and load. Equation (6) limits curtailment to the available PV generation, while (7) prevents load shedding from exceeding demand.

$$P^{DG,min}u^{DG}_{y,d,t} \le P^{DG}_{y,d,t} \tag{8}$$

$$P^{DG}_{y,d,t} \le M^{DG}u^{DG}_{y,d,t} \tag{9}$$

$$P^{DG}_{y,d,t} \le S^{DG} \tag{10}$$

Equations (8) and (9) link DG output to the binary commitment state, while (10) limits output by the installed capacity. The case study assumes zero minimum generation and zero no-load cost, although the general terms are retained for compatibility with other DG technologies.

$$0 \le P^{imp}_{y,d,t} \le P^{tie} u^{imp}_{y,d,t} \tag{11}$$

$$0 \le P^{exp}_{y,d,t} \le P^{tie} u^{exp}_{y,d,t} \tag{12}$$

$$u^{imp}_{y,d,t} + u^{exp}_{y,d,t} \le 1 \tag{13}$$

$$C^{exp}_{y,d,t} = (1 - \beta^{exp}) C^{imp}_{y,d,t} \tag{14}$$

$$P^{ch}_{y,d,t} + P^{exp}_{y,d,t} \le P^{imp}_{y,d,t} + \phi^{PV}_y P^{PV,cf}_{d,t} S^{PV} - P^{PV,curt}_{y,d,t} + P^{dch}_{y,d,t} \tag{15}$$

Equations (11) and (12) restrict grid transactions to the tie-line rating, while (13) prevents simultaneous import and export. Equation (14) defines the export price as a discounted fraction of the import price, where $\beta^{exp}$ is the export-price discount, set to 0.20 in this study. Equation (15) ensures that aggregate BESS charging and grid export are supported only by grid imports, net PV generation, and BESS discharge. DG may operate simultaneously to serve local demand, but it does not contribute to the charging and export energy balance.

$$S^{B,av}_y = H_y S^B \tag{16}$$

$$SOC^{min} H_y S^B \le E^B_{y,d,t} \le SOC^{max} H_y S^B \tag{17}$$

$$E^{B,init}_y = SOC^{init} H_y S^B \tag{18}$$

Equation (16) converts the installed nameplate capacity $S^B$ into the available energy capacity $S^{B,av}_y$ using BESS SOH $H_y$. Equations (17) and (18) define the allowable stored-energy range and the initial stored energy for each year.

$$u^{ch}_{y,d,t} + u^{dch}_{y,d,t} \le 1 \tag{19}$$

$$0 \le P^{ch}_{y,d,t} \le M^B u^{ch}_{y,d,t} \tag{20}$$

$$P^{ch}_{y,d,t} \le \frac{S^B}{\tau^{ch}} \tag{21}$$

$$0 \le P^{dch}_{y,d,t} \le M^B u^{dch}_{y,d,t} \tag{22}$$

$$P^{dch}_{y,d,t} \le \frac{S^B}{\tau^{dch}} \tag{23}$$

Equation (19) prohibits simultaneous charging and discharging. Equations (20) and (22) link BESS power to the corresponding binary operating states, while (21) and (23) define the charging and discharging power ratings from the nameplate energy capacity and rated durations.

$$\eta^{ch}_y = \eta^{dch}_y = \sqrt{\eta^{RTE}_y} \tag{24}$$

$$E^B_{y,d,t} = E^B_{prev} + \left( \eta^{ch}_y P^{ch}_{y,d,t} - \frac{P^{dch}_{y,d,t}}{\eta^{dch}_y} \right) \Delta t \tag{25}$$

$$E^B_{y,|D|,|T|} = E^{B,init}_y \ \forall y \in Y \tag{26}$$

Equation (24) allocates the specified RTE symmetrically between charging and discharging. Equation (25) maintains chronological SOC continuity across all hours and days, where $E^B_{prev}$ is the stored energy in the preceding interval, while (26) prevents artificial energy transfer between planning years. During the planning stage, $H_y$ and $\eta^{RTE}_y$ are fixed at their initial battery values. This modeling stage therefore captures load growth and PV degradation but does not anticipate usage-dependent BESS capacity fade or RTE decline.

### *B. Sequential Lifecycle Validation*

After Stage 1, $S^{DG}$, $S^{PV}$, and $S^B$ are fixed, and a one-year operating model is solved sequentially for each lifecycle year. At the beginning of year $y$, the model is updated using the corresponding load level, PV performance factor, BESS SOH, RTE, available energy capacity, and tariff profile. Battery condition is held constant within each annual optimization. The resulting chronological SOC trajectory is then used to calculate degradation, which is applied at the beginning of the following year. The normalized SOC trajectory is defined as

$$SOC_{y,\tau} = \frac{E^B_{y,\tau}}{H^{start}_y S^B} \tag{27}$$

where $\tau$ denotes the chronological hourly index and runs continuously from 1 to 8,760 across the whole year rather than resetting each day, so that cycles spanning midnight are preserved. A rainflow algorithm following ASTM E1049-85, the standard practice for cycle counting in fatigue analysis [33], extracts cycles $k$ with depth $D_{y,k}$ and count $n_{y,k}$, where complete and residual half-cycles contribute 1 and 0.5, respectively. The annual equivalent full cycles ($EFC_y$) are thus calculated as

$$EFC_y = \sum_k D_{y,k}\, n_{y,k} \tag{28}$$

Equation (28) is reported as an interpretable battery-utilization metric. Degradation is instead calculated from the depth-dependent damage associated with each identified cycle. Let $N\left(D_{y,k}\right)$ denote the cycle life at cycle depth $D_{y,k}$. Values between the available depth-of-discharge (DOD) points are obtained through linear interpolation. For cycle depths below the minimum tabulated value, the cycle life at the minimum available depth is used. The annual fraction of cycle life consumed is calculated using Miner's rule [38]:

$$\lambda_y = \sum_k \frac{n_{y,k}}{N(D_{y,k})} \tag{29}$$

where $\lambda_y$ is the dimensionless fraction of the adopted cycle-life budget consumed in year $y$. The adopted DOD-dependent cycle-life curve is mapped to a 20-percentage-point SOH loss for planning purposes as

$$\Delta H_y = (1 - H^{curve,EOL}) \lambda_y \tag{30}$$

where $H^{curve,EOL} = 0.8$. The end-of-year SOH $H^{end}_y$ and corresponding available energy capacity $S^{B,av,end}_y$ are

$$H^{end}_y = \max(0, H^{start}_y - \Delta H_y) \tag{31}$$

$$S^{B,av,end}_y = H^{end}_y S^B \tag{32}$$

Equations (29)–(32) update battery condition from the annual cycling history. Calendar aging is not included; therefore, the modeled SOH reduction represents cycling-induced degradation under the adopted cycle-life curve. The RTE is updated as a function of the start-of-year SOH $H^{start}_y$:

$$\eta^{RTE}_y = \min\{1, \max[0, a_\eta H^{start}_y + b_\eta]\} \tag{33}$$

$$a_\eta = 0.5,\ b_\eta = 0.4 \tag{34}$$

where $a_\eta$ and $b_\eta$ are the slope and intercept of the adopted linear SOH–RTE mapping. Equation (33) is a two-point planning approximation anchored at the assumed new-BESS condition (H = 1.00, RTE = 0.90) and the SLB entry condition (H = 0.80, RTE = 0.80), consistent with the reported association between battery aging and energy-efficiency decline [12], [16]. It is not a universal fitted law; therefore, the ablation shares in Section V-C are conditional on it.

$$\phi^{PV}_{y+1} = \phi^{PV}_y (1 - \delta^{PV}) \tag{35}$$

Equation (35) is the sequential equivalent of the closed-form relationship in (3). Let $H^{init}$ denote the battery-type-specific initial SOH. The operational EOL is defined as

$$H^{op,EOL} = H^{init} - 0.20 \tag{36}$$

A replacement is triggered when

$$H_y^{end} \leq H^{op,EOL} \tag{37}$$

provided that $y$ is not the final year, ensuring no replacement is installed after the planning horizon. The resulting EOL thresholds are 0.80 for the new BESS and 0.60 for the SLB. A replacement triggered after year $y$ is installed before year $y+1$ and restores the same battery type to its specified initial SOH and RTE.

$$C_y^{rep} = \rho_B^{rep} C_B^{cap} S^B \tag{38}$$

$$C_y^{rep,pv} = \frac{C_y^{rep}}{(1+r)^y} \tag{39}$$

Equation (38) prices replacement as the fraction $\rho_B^{rep}$, set to 0.80, of the applicable installed battery cost. Equation (39) discounts this expenditure by $y$ years because replacement occurs at the beginning of year $y+1$.

$$\begin{aligned} NPC^{val} = {} & C_{DG}^{cap} S^{DG} + C_{PV}^{cap} S^{PV} + C_B^{cap} S^B \\ & + \textstyle\sum_{y=1}^{Y} \frac{C_y^{op,econ}}{(1+r)^{y-1}} + \sum_{y=1}^{Y-1} z_y^{rep} C_y^{rep,pv} \end{aligned} \tag{40}$$

Equation (40) represents the validated economic NPC and it combines initial capital investment, discounted annual operating expenditure, and discounted battery-replacement expenditure. The binary indicator $z_y^{rep}$ equals one when (37) triggers replacement after year $y$, and zero otherwise. $C_y^{op,econ}$ includes DG operating cost, DG and PV fixed O&M, grid-import expenditure, and grid-export revenue. The load-shedding penalty is excluded from $NPC^{val}$, and decommissioning and salvage/residual values are not modeled.

Annual and cumulative lifecycle ENS are calculated as

$$ENS_y = \textstyle\sum_{d \in D} \sum_{t \in T} P_{y,d,t}^{LS} \Delta t \tag{41}$$

$$ENS^{life} = \textstyle\sum_{y \in Y} ENS_y \tag{42}$$

Equations (41) and (42) convert hourly unserved power to annual ENS and sum it over the lifecycle. The validation also records the first and last ENS years, outage duration, event count, maximum annual ENS, and maximum load shed. A candidate is classified as lifecycle reliable when

$$ENS_y \leq \varepsilon^{ENS} \; \forall y \in Y \tag{43}$$

$$P_{y,d,t}^{LS} \leq \varepsilon^{LS} \; \forall y, d, t \tag{44}$$

The adopted tolerances are $\varepsilon^{ENS} = 10^{-4}$ MWh/year and $\varepsilon^{LS} = 10^{-5}$ MW, distinguishing numerical residuals from material adequacy violations. Because load, PV availability, tariffs, and tie-line availability are deterministic, ENS is a chronological adequacy metric rather than a probabilistic reliability expectation.

### C. *Reliability-Guided Investment Refinement*

If the initial portfolio fails either (43) or (44), DAVIR-MG evaluates separate BESS, DG, and PV augmentation pathways. Each pathway changes one capacity while holding the other two at their original values, and every candidate undergoes the full sequential lifecycle validation:

$$S^{B,new} = S^{B,0} + \Delta S^B \tag{45}$$

$$S^{DG,new} = S^{DG,0} + \Delta S^{DG} \tag{46}$$

$$S^{PV,new} = S^{PV,0} + \Delta S^{PV} \tag{47}$$

The superscript $'0'$ denotes the original planning portfolio. For BESS and DG, bisection is applied between an infeasible lower bound and a reliable upper bound until the interval falls below the selected resolution; the final capacity is then rounded upward and revalidated. PV is screened at predefined capacities because additional PV may not reduce ENS when shortage hours have low solar availability or curtailment limits charging opportunities.

Let $\mathcal{F}$ denote the set of evaluated candidates satisfying (43) and (44). The preferred economic alternative is

$$j^* = \arg\min_{j \in \mathcal{F}} NPC_j^{val} \tag{48}$$

Equation (48) selects the candidate with the lowest validated economic NPC within $\mathcal{F}$. The load-shedding penalty is excluded from the ranking, while SOH, RTE, replacement timing, EFC, and operating behavior distinguish alternatives with similar cost but different technical outcomes. Thus, $j^*$ denotes the least-cost feasible candidate within the evaluated technology-specific refinement set.

### D. *Conditional Cost-ENS and SLB Break-Even Analyses*

Complete restoration is defined as the BESS capacity that eliminates ENS; partial restoration is any intermediate capacity that reduces ENS without eliminating it. To examine the trade-off between them, BESS capacity is evaluated at a sequence of values between the original portfolio and the fully reliable BESS-refined solution, with DG and PV held fixed. The resulting frontier is conditional on those fixed capacities. For consecutive tested capacities $i$ and $i+1$, the marginal lifecycle cost per MWh of cumulative ENS removed is

$$MC_{i \to i+1}^{ENS} = \frac{NPC_{i+1}^{val} - NPC_i^{val}}{ENS_i^{life} - ENS_{i+1}^{life}} \tag{49}$$

Equation (49) is a conditional cost-effectiveness metric, not a value of lost load, because interrupted service is not monetized. A tested candidate is nondominated within the evaluated set when no other tested candidate has both a lower or equal validated NPC and a lower or equal cumulative ENS, with at least one strict improvement. The result is therefore a conditional BESS-refinement Cost-ENS frontier at fixed DG and PV capacity, not a global system-planning frontier.

The SLB break-even analysis is conducted only after new and second-life batteries have been separately sized to satisfy the same lifecycle-reliability criterion. DG capacity, PV capacity, tariff structure, PV degradation, load growth, and other non-storage assumptions are held fixed. Let $f^{SLB}$ denote the SLB installed-cost fraction relative to the new-BESS cost. The corresponding SLB discount is

$$\beta^{SLB} = 1 - f^{SLB} \tag{50}$$

The break-even installed-cost fraction $f^{SLB,*}$ satisfies

$$NPC_{SLB}^{val}(f^{SLB,*}) = NPC_{new}^{val} \tag{51}$$

Two replacement-pricing conventions are considered, reflecting whether cheap second-life cells can be obtained years later, which is uncertain because the second-life cost advantage is projected to narrow as new-battery costs fall [39]. Repeat-SLB pricing applies the initial SLB discount to future replacements, whereas new-cost pricing charges replacements at the full new-BESS installed cost. Both conventions restore the same SLB technical condition. Once the reliable new-BESS and SLB capacities are fixed, only the cost fraction is varied: capacities, degradation trajectories, replacement timing, and dispatch are not recomputed at each price point, which isolates the effect of price on the comparison.

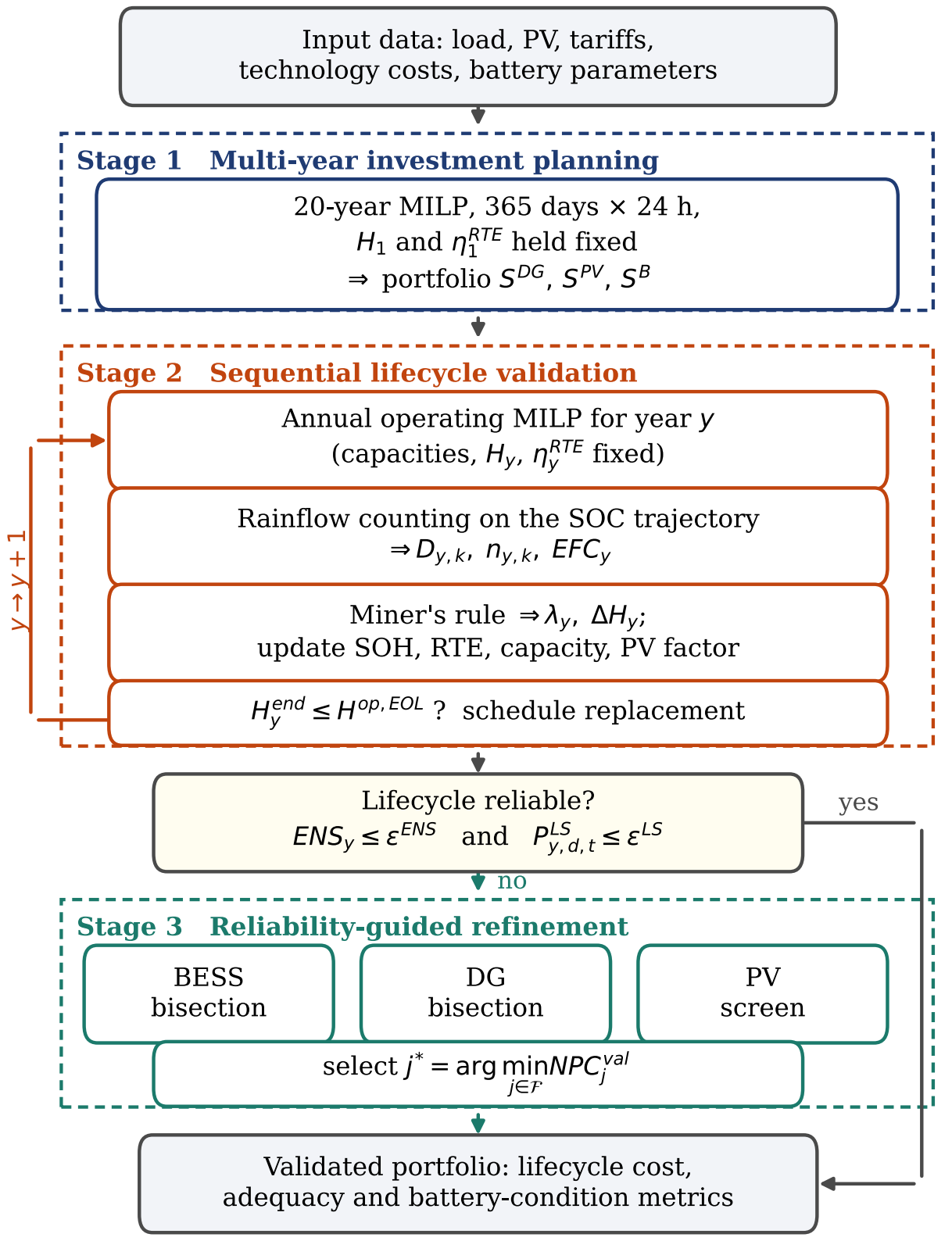

Fig. 1. Structure of the DAVIR-MG framework.

## IV. Case Studies and Experimental Design

### A. Test System and Input Data

DAVIR-MG is evaluated on a grid-connected residential MG in Houston, Texas, comprising a natural-gas DG, a PV system, a lithium iron phosphate BESS, and a 0.10-MW bidirectional grid connection (Fig. 2). The restrictive tie line represents a grid-constrained interconnection, so the adequacy findings apply to similarly constrained MGs rather than all grid strengths.

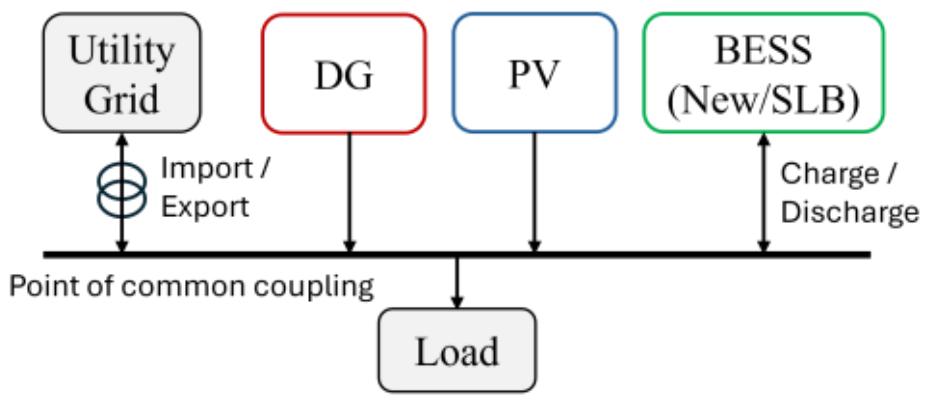

Fig. 2. Grid-connected residential test microgrid.

Hourly residential demand is obtained from OpenEI [40] and grows by 0.5% annually; year-one demand is 1,514.6 MWh with a 0.8 MW peak. PV capacity factors are generated using PVWatts [41]. Fixed, time-of-use (TOU), and ERCOT Houston day-ahead price structures are considered [42]–[44], with exports compensated at 80% of the import price. All cases retain 8,760 chronological hours per year to preserve tariff response and rainflow cycle identification.

### B. Techno-Economic Assumptions

Table I summarizes the planning, cost, and battery assumptions in real U.S. dollars discounted at 3%. DG and PV costs follow contemporary technology-cost sources [45]-[47]. The new-BESS installed cost of \$0.476M/MWh is the reference against which the SLB installed-cost fraction is defined. The moderate and high discounts, 30% and 70%, correspond to the range of second-life cost advantage projected in [39] and are used as scenario endpoints rather than forecasts for a particular market. Both battery types are assigned a 20-percentage-point SOH window between installation and replacement, so the new BESS operates from 1.00 to 0.80 and the SLB from 0.80 to 0.60. The same manufacturer-reported LFP cycle-life curve [48] is applied to both batteries with rainflow counting and Miner's rule, enabling a controlled comparison but not capturing the heterogeneity of repurposed battery aging. Replacement restores each battery type to its initial SOH and RTE.

TABLE I
Planning, Cost, and Battery Assumptions

| Parameter | Value |
| --- | --- |
| **Planning** | |
| Horizon, operating days, resolution | 20 years, 365 days, 1 hour |
| Annual load growth, discount rate | 0.5%, 3% |
| Tie-line capacity, export price | 0.10 MW, 80% of import price |
| PV degradation, base (sensitivities) | 1.0% (0.5%, 1.5%) per year |
| **Costs** | |
| DG capital, operating cost, fixed O&M | \$1.15M/MW, \$60/MWh, \$40k/MW-y |
| PV capital, fixed O&M | \$1.00M/MW, \$19k/MW-y |
| New-BESS installed cost | \$0.476M/MWh |
| Replacement-cost fraction | 80% of applicable installed cost |
| SLB installed-cost fraction | 0.70 (moderate), 0.30 (high) |
| **Battery condition (new BESS / SLB)** | |
| Initial SOH | 1.00 / 0.80 |
| Initial RTE | 0.90 / 0.80 |
| Operational EOL SOH | 0.80 / 0.60 |
| Charge/discharge time, SOC range | 1 h / 1 h, 0 to 100% |

### C. Planning Scenario Matrix

Table II lists nine scenario labels corresponding to seven unique optimization cases, since P2 is identical to T1 and B1 to T3. Together they span tariff structure, PV degradation rate, and battery condition. Cases T1 to T3 isolate the pricing structure for a new BESS at the base PV degradation rate. Cases P1-P3 form the PV degradation sensitivity under the fixed tariff. Cases B1-B3 compare the new BESS against moderate and high discount SLBs under the wholesale tariff. Each case is solved with the multi-year planning model and then validated over 20 sequential years. Planning-stage and validated NPC are reported separately, the former reflecting the degradation-naive investment problem and the latter including degradation-dependent operation and discounted replacement, and the load-shedding penalty is excluded from validated NPC.

TABLE II
PLANNING AND LIFECYCLE-VALIDATION CASES

| Case | Tariff | Battery | SOH$_0$ | RTE$_0$ | Cost frac. | PV deg. |
|---|---|---|---|---|---|---|
| T1 | Fixed | New | 1.00 | 0.90 | 1.00 | 1.0% |
| T2 | TOU | New | 1.00 | 0.90 | 1.00 | 1.0% |
| T3 | Wholesale | New | 1.00 | 0.90 | 1.00 | 1.0% |
| P1 | Fixed | New | 1.00 | 0.90 | 1.00 | 0.5% |
| P2[a] | Fixed | New | 1.00 | 0.90 | 1.00 | 1.0% |
| P3 | Fixed | New | 1.00 | 0.90 | 1.00 | 1.5% |
| B1[b] | Wholesale | New | 1.00 | 0.90 | 1.00 | 1.0% |
| B2 | Wholesale | SLB | 0.80 | 0.80 | 0.70 | 1.0% |
| B3 | Wholesale | SLB | 0.80 | 0.80 | 0.30 | 1.0% |

*[a] P2 is identical to T1 and is listed to show the PV degradation sensitivity.*
*[b] B1 is identical to T3 and is listed only to complete the BESS type sensitivity.*

### D. Analysis Sequence and Implementation

Section V reports six analyses in sequence: planning and lifecycle validation cases; selection of a degradation-critical case; degradation-mechanism ablation on the selected portfolio; technology-specific reliability refinement; a conditional BESS Cost-ENS assessment; and an equal-reliability SLB break-even analysis. Bisection resolutions are 0.01 MWh for BESS and 0.01 MW for DG, with the final candidate rounded upward to the next increment and revalidated; PV is evaluated through a capacity screen rather than bisection. The models are implemented in Pyomo and solved with Gurobi 12.0.3, each planning and annual validation problem being a MILP. Across all cases, the multi-year planning model and the sequential lifecycle validation each solve in under 300 s on an Intel Xeon W-2195 CPU at 2.30 GHz with 128 GB RAM. Every reported case is additionally checked for solver status, power-balance and terminal-SOC residuals, operating-state consistency, and agreement between annual and summary-level economics.

## V. RESULTS AND DISCUSSION

### A. Initial Planning and Lifecycle Validation

All seven unique cases were solved to optimality and validated over the full horizon. Table III reports the planning capacities, both NPC measures, and the principal reliability and battery-life indicators. Validated NPC exceeds planning NPC in every case, by 3.0% to 11.6%. The planning stage assumes the battery never ages, so it omits both the cost of operating a degraded battery and the cost of buying a replacement; validation adds both. Although cumulative ENS is small relative to 31,774 MWh of lifecycle demand, it exceeds the prescribed tolerance and first appears only in years 14–17, so an initial-year adequacy check would miss it.

#### 1) Effect of tariff structure:

DG capacity remains near 0.55 MW across T1–T3, while PV and BESS respond strongly to the tariff structure. The fixed and TOU cases select roughly twice the PV capacity of the wholesale case and slightly less storage. T2 has the lowest validated NPC but the highest ENS and battery utilization, at about 7,289 lifecycle EFC against 6,142 for the fixed tariff and 6,064 for wholesale, with replacements at the start of years 8 and 15 rather than 9 and 18. Lower cost therefore does not imply lower battery stress or better late-life adequacy.

TABLE III
INITIAL PLANNING AND LIFECYCLE-VALIDATION RESULTS

| Case | DG (MW) | PV (MW) | BESS (MWh) | Plan. NPC ($M) | Valid. NPC ($M) | ENS (MWh) | 1st ENS yr | Repl. yr(s) |
|---|---|---|---|---|---|---|---|---|
| T1 | 0.5523 | 0.4801 | 0.4363 | 2.3147 | 2.5615 | 0.1842 | 15 | 9, 18 |
| T2 | 0.5524 | 0.4444 | 0.4363 | 2.2734 | 2.5367 | 0.2760 | 14 | 8, 15 |
| T3 | 0.5536 | 0.2220 | 0.4834 | 2.3233 | 2.5828 | 0.1521 | 16 | 10, 19 |
| P1 | 0.5521 | 0.4801 | 0.4363 | 2.2877 | 2.5374 | 0.1646 | 15 | 9, 17 |
| P3 | 0.5524 | 0.4881 | 0.4363 | 2.3418 | 2.5882 | 0.1770 | 15 | 9, 18 |
| B2 | 0.5503 | 0.2153 | 0.6613 | 2.3212 | 2.4624 | 0.5581 | 16 | 11 |
| B3 | 0.5495 | 0.1704 | 0.6797 | 2.1932 | 2.2591 | 0.4209 | 17 | 12 |

*Cases P2 and B1 are identical to T1 and T3 and are not repeated.*
*Replacement years denote the year at whose start the replacement is installed.*

#### 2) Effect of PV degradation:

Under the fixed tariff, validated NPC rises monotonically with the assumed PV degradation rate, from \$2.5374 million in P1 to \$2.5882 million in P3. Installed PV capacity does not: P1 and P2 both select about 0.480 MW while P3 selects 0.488 MW, and cumulative ENS rises from P1 to P2 before falling slightly in P3. Faster PV degradation therefore increases lifecycle cost, while the optimizer can offset lost output through marginally more PV or greater reliance on DG and storage.

#### 3) Initial comparison of new and second-life batteries:

Under the wholesale tariff, the new BESS is sized at 0.483 MWh, while the SLB cases require 0.661 and 0.680 MWh because of their lower initial SOH and RTE. Case B2 has a planning NPC close to T3 but a lower validated NPC, and B3 reduces costs further. The SLB cases, however, incur more cumulative ENS than T3. The deep-discount case B3 is the most striking: a 70% lower battery price lowers planning NPC by 5.6% and validated NPC by 12.5% relative to T3. These scenario results still cannot establish SLB superiority, because the portfolios do not deliver equal reliability. That comparison is deferred to Section V-F.

### B. Selection of the Degradation-Critical Case

Case B2 is selected because it combines a moderate 30% SLB discount with the largest cumulative ENS. Fig. 3 illustrates the lifecycle mechanism in detail.

The SLB falls from an initial SOH of 0.80 to 0.595 by the end of year 10, crossing the 0.60 operational EOL threshold and triggering replacement at the start of year 11. The replacement battery then degrades to 0.607 by the end of year 20 without reaching a second replacement. ENS first appears in year 16 and grows monotonically to 0.263 MWh in year 20, while outage activity rises from one event and one hour to three events and six hours. The shortfall is progressive rather than an isolated numerical violation, and it does not arise from intensified cycling: annual EFC declines steadily from 265.8 in year 1 to 238.5 in year 20, since a degraded battery moves less energy. The operative mechanism is shrinking usable capacity combined with compounding demand growth, which narrows the operating margin during hours when the 0.1 MW tie line and the DG are simultaneously binding. The annual peak of 0.800 MW already exceeds their combined firm capacity of 0.650 MW, so the BESS must contribute during those hours. That condition arises in only 10 hours of the year, predominantly winter mornings with negligible PV output, which is why degradation-induced shortfalls emerge late in the horizon rather than early.

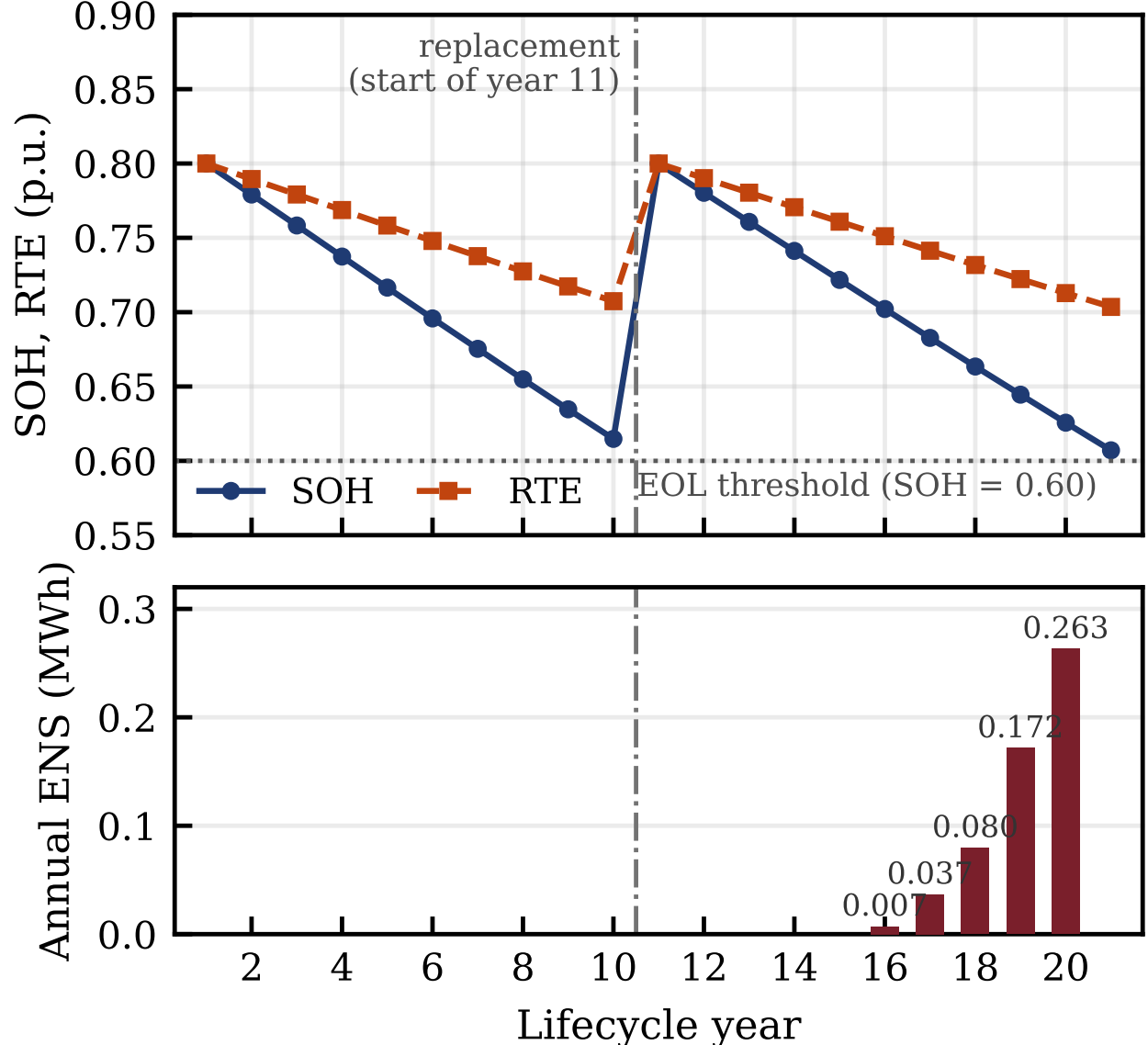

Fig. 3. B2 lifecycle SOH, RTE, and annual ENS.

## C. Degradation-Mechanism Ablation

Table IV reports the four ablation cases evaluated on the fixed B2 capacities, and Fig. 4 decomposes the resulting cumulative ENS. Case A0 produces no ENS when capacity and RTE are held at their initial SLB values. This closure test confirms that the shortfall in A3 originates in the modeled degradation mechanisms rather than in the annual validation sequence. As depicted, capacity fade dominates. Case A1 alone produces 0.4261 MWh, or 76.4% of the full A3 impact, because declining usable capacity withdraws stored energy precisely during hours when DG and import limits bind. RTE fade alone produces 0.0598 MWh, or 10.7%: smaller, but not negligible, since lower efficiency raises charging losses and reduces net energy returned on discharge. Lower RTE alters the charging and discharging trajectory while reduced capacity alters the energy range across which those losses accrue, so the joint impact exceeds the sum of the isolated effects.

TABLE IV
DEGRADATION-MECHANISM ABLATION RESULTS

| Case | Mechanism | Valid. NPC ($M) | ENS (MWh) | 1st ENS yr | Outage hours | EFC | Repl. yr |
|---|---|---|---|---|---|---|---|
| A0 | None | 2.3212 | 0.0000 | – | 0 | 5,048 | – |
| A1 | Capacity fade | 2.4558 | 0.4261 | 17 | 14 | 5,470 | 11 |
| A2 | RTE fade | 2.4556 | 0.0598 | 19 | 4 | 4,703 | 12 |
| A3 | Capacity + RTE | 2.4624 | 0.5581 | 16 | 18 | 5,072 | 11 |

*A2 holds usable capacity fixed while the internal state drives RTE and replacement.*

Case A2 is a counterfactual mechanism-isolation case in which usable capacity is fixed while the internal damage state continues to determine RTE and replacement. It is therefore not a physically complete aging representation. Capacity fade alone accounts for 76.4% of A3 ENS, RTE decline alone for 10.7%, and their interaction for the remaining 12.9%.

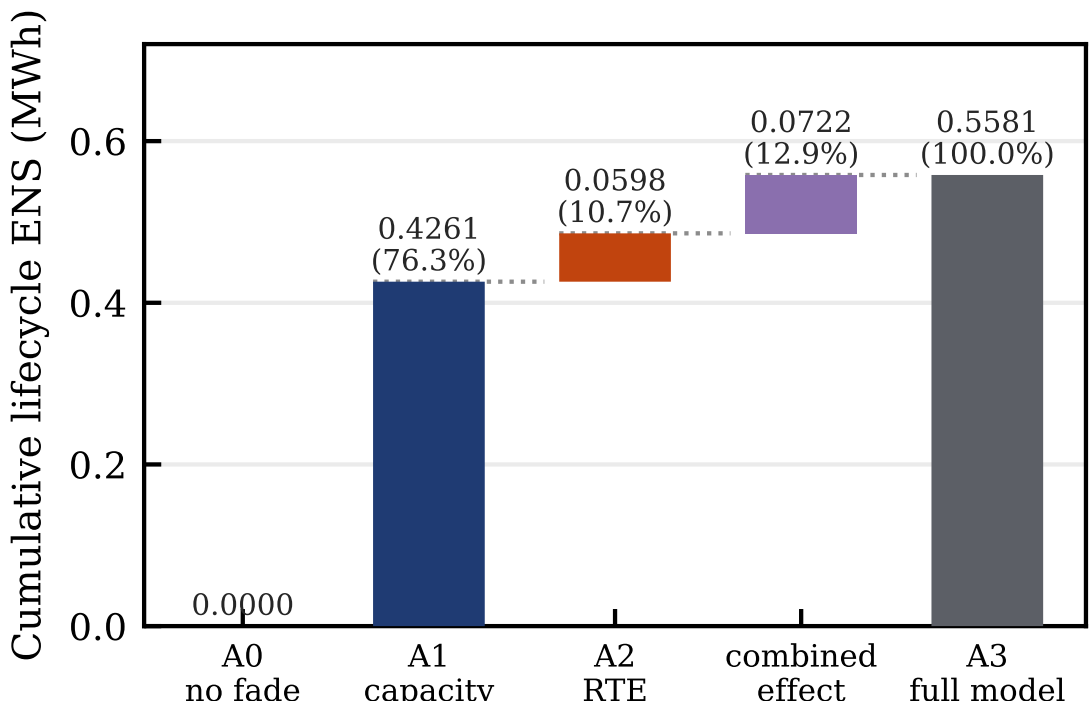

Fig. 4. Decomposition of lifecycle ENS in B2.

## D. Technology-Specific Reliability Refinement

Table V summarizes the three refinement pathways for the B2 baseline, denoted R0. The original 0.6613-MWh BESS is lifecycle infeasible, while bisection and upward rounding identify 0.82 MWh as reliable. The 0.159-MWh addition increases initial investment by $52,951 and validated NPC to $2.5324 million while eliminating ENS. It also lowers battery stress: lifecycle EFC falls from 5,072 to 4,355, replacement shifts from year 11 to 13, and final SOH and RTE improve from 0.607 and 0.704 to 0.671 and 0.736.

DG refinement brackets 0.58 MW as infeasible and 0.59 MW as reliable, adding about 0.040 MW for a validated NPC of $2.5323 million. Its $102 advantage over R1 is immaterial at the precision of the cost inputs, so R1 and R2 are a practical economic tie. The pathways differ technically: R2 adds only about 1.1 MWh of DG generation over 20 years but leaves battery replacement and final condition unchanged, whereas R1 reduces cycling and improves battery health. A planner prioritizing battery condition may prefer R1, while one seeking to avoid storage expansion may prefer R2; reporting both is more informative than treating storage augmentation as the default correction. The preferred pathway may also change when additional objectives, such as emissions, fuel security, or technology-policy constraints, are explicitly modeled.

PV refinement does not restore adequacy in the five screened cases up to 5.0 MW. The closest case, 1.7225 MW, removes 99.9% of ENS but leaves a year-17 residual above tolerance, because the binding winter-morning hours have negligible PV output. It is also uneconomic: capital and discounted replacement costs alone total $2.85 million, exceeding the entire validated NPC of R1, while lifecycle EFC rises 41% and a second replacement becomes necessary. The adequacy result is specific to the screened capacities, but the cost gap widens monotonically with PV size, so the economic conclusion holds across the range. This outcome supports evaluating refinement pathways separately, since the timing of the binding hours determines which resource can restore adequacy at all.

TABLE V
TECHNOLOGY-SPECIFIC REFINEMENT RESULTS

| Case | Refine | DG (MW) | PV (MW) | BESS (MWh) | Valid. NPC ($M) | ENS (MWh) | Repl. yr |
|---|---|---|---|---|---|---|---|
| R0 | None | 0.5503 | 0.2153 | 0.6613 | 2.4624 | 0.5581 | 11 |
| R1 | BESS | 0.5503 | 0.2153 | 0.8200 | 2.5324 | 0.0000 | 13 |
| R2 | DG | 0.5900 | 0.2153 | 0.6613 | 2.5323 | 0.0000 | 11 |
| R3 | PV | 0.5503 | 1.7225[a] | 0.6613 | >2.850[b] | 0.00039 | 7,13 |

[a] *Closest screened PV capacity; no screened capacity satisfies the criterion.*
[b] *Lower bound based on capital & replacement costs; full validated NPC is higher.*

### *E. Conditional BESS Cost-ENS Analysis*

Fig. 5 shows the conditional frontier obtained by varying BESS capacity between the R0 value and the reliable R1 value while DG and PV remain fixed. All five candidates are nondominated, since each capacity increase raises validated NPC and lowers cumulative ENS.

The first increment provides most of the reliability gain: increasing BESS capacity from 0.661 to 0.700 MWh, only 24.4% of the full augmentation, removes 46.8% of ENS at $57,606 per MWh of ENS removed. At 0.780 MWh, 92.8% is removed; the final 0.040-MWh increment eliminates the remaining 7.2% at $491,187/MWh. Full restoration raises validated NPC by 2.85%, comprising $52,951 in initial BESS cost and $22,169 in discounted replacement cost, partly offset by $5,077 in operating savings. It also reduces lifecycle EFC by 14.1%, delays replacement two years, and raises final available capacity from 0.401 to 0.550 MWh. The intermediate points therefore expose budget-constrained trade-offs hidden by a binary feasibility test.

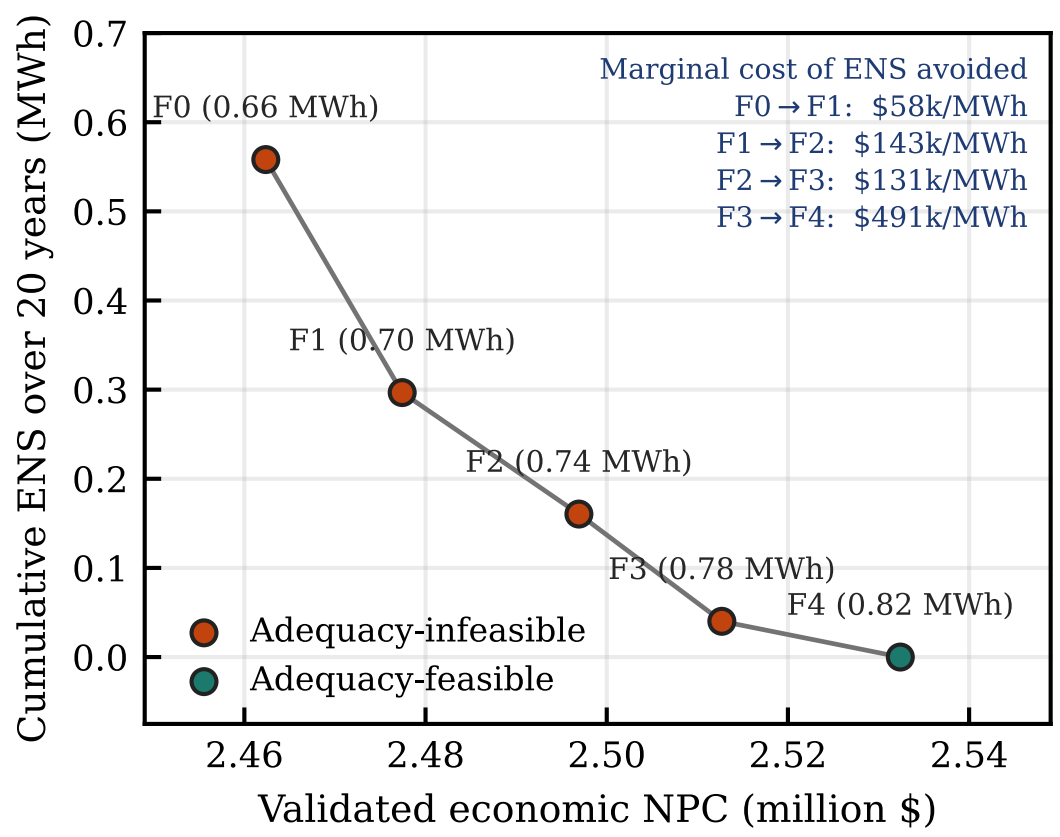

Fig. 5. Conditional Cost-ENS frontier at fixed DG and PV capacities.

### *F. Equal-Reliability SLB Break-Even Analysis*

T3, B2, and B3 are not directly comparable because they provide different reliability. The new BESS and SLB are therefore separately sized to satisfy the same criteria under identical non-storage assumptions. Bisection gives 0.64 MWh for the new BESS and 0.82 MWh for the SLB (Table VI). Although the SLB requires 28.1% more nameplate capacity, its 0.80 initial SOH yields 0.656 MWh of usable energy versus 0.640 MWh for the new BESS, only 2.5% more; most of the nameplate premium therefore compensates for lower initial SOH, with the remainder offsetting lower RTE.

The reliable new-BESS portfolio has a validated NPC of $2.5760 million. Under repeat-SLB replacement pricing, cost parity occurs at a 77.14% cost fraction, requiring a 22.86% discount; pricing replacement at the new-BESS cost lowers the break-even fraction to 64.31%, requiring a 35.69% discount (Fig. 6). At a 30% discount, the SLB saves $43,580 (1.69%) under repeat pricing but costs $22,225 more (0.86%) under new-cost pricing. At a 70% discount, it remains cheaper under both conventions, by $287,691 (11.17%) and $134,145 (5.21%). Thus, SLB competitiveness depends on replacement pricing as well as acquisition cost.

TABLE VI
EQUAL-RELIABILITY NEW-BESS AND SLB PORTFOLIOS

| Attribute | New BESS | SLB |
|---|---|---|
| Nameplate capacity | 0.64 MWh | 0.82 MWh |
| Initial SOH | 1.00 | 0.80 |
| Initial usable capacity | 0.640 MWh | 0.656 MWh |
| Initial RTE | 0.90 | 0.80 |
| Lifecycle ENS | 0 MWh | 0 MWh |
| First replacement year | 11 | 13 |
| Lifecycle EFC | 5,090 | 4,355 |
| Final SOH | 0.808 | 0.671 |
| Final RTE | 0.804 | 0.736 |
| New-BESS reference NPC | $2.5760M | - |
| SLB NPC, repeat-SLB pricing (30% discount) | - | $2.5324M |
| SLB NPC, new-cost replacement (30% discount) | - | $2.5982M |

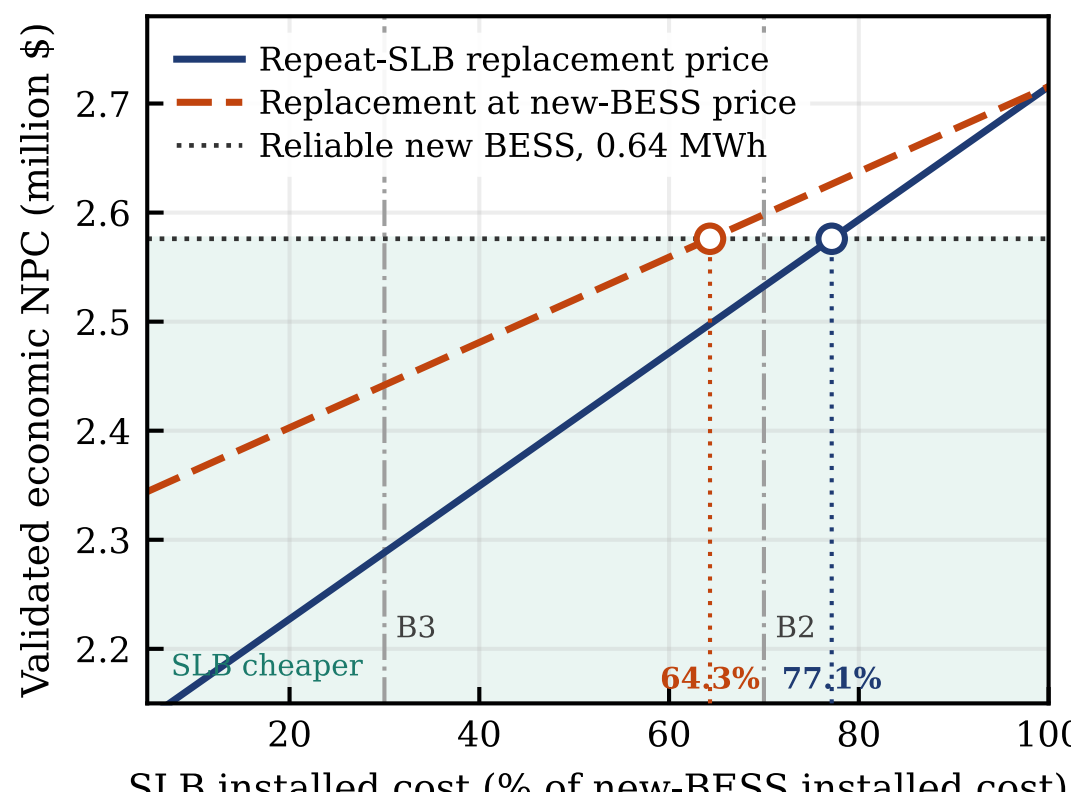

Fig. 6. Equal-reliability SLB break-even under two replacement-pricing conventions.

## VI. CONCLUSION

This study developed DAVIR-MG, a three-stage framework linking multi-year investment planning with sequential degradation-aware validation and reliability-guided refinement. Across seven unique cases, validated NPC exceeded degradation-naive planning NPC by 3.0%–11.6%, while adequacy shortfalls emerged only after year 13, demonstrating that initially feasible portfolios may become unreliable later in life. In the selected SLB case, capacity fade alone accounted for 76.4% of cumulative ENS, RTE decline alone for 10.7%, and their interaction for the remaining 12.9%. BESS and DG refinements restored adequacy at nearly identical lifecycle cost but through different technical pathways, whereas the conditional Cost-ENS analysis showed that partial storage augmentation can remove most ENS at substantially lower marginal cost than complete restoration.

Under equal-reliability sizing, the minimum SLB discount required for cost parity increased from 22.86% to 35.69% when replacement was priced at the new-BESS cost, highlighting the importance of future replacement assumptions in SLB valuation. Overall, the results show that degradation-aware validation can materially alter both reliability and economic conclusions and that targeted refinement provides more informative investment choices than degradation-naive sizing alone. These case-specific findings assume a grid-constrained single-bus MG, deterministic annual profiles, cycle-only aging, annual condition updates, a common cycle-life curve, and a two-point SOH–RTE mapping. Future work will examine interconnection strength, interannual variability, alternative SOH–RTE relationships, finer degradation updates, and network-constrained stochastic planning.